\documentclass[journal]{IEEEtran}
\usepackage{cite}
\usepackage{amsmath,amssymb,amsfonts}
\usepackage{cuted}
\usepackage{graphicx}
\usepackage{textcomp,nicefrac}
\usepackage{wrapfig}
\usepackage{soul}
\usepackage{listings}
\usepackage{color}
\usepackage{floatrow}
\usepackage{xcolor}
\usepackage{hyperref}
\usepackage{tabularx}
\usepackage{graphicx}
\usepackage{comment}
\usepackage{wasysym}
\usepackage[capitalize]{acro}
\usepackage{threeparttable}
\usepackage{lipsum}

\usepackage{scalerel}
\usepackage{tikz}
\usetikzlibrary{svg.path}
\definecolor{orcidlogocol}{HTML}{A6CE39}
\tikzset{
  orcidlogo/.pic={
    \fill[orcidlogocol] svg{M256,128c0,70.7-57.3,128-128,128C57.3,256,0,198.7,0,128C0,57.3,57.3,0,128,0C198.7,0,256,57.3,256,128z};
    \fill[white] svg{M86.3,186.2H70.9V79.1h15.4v48.4V186.2z}
                 svg{M108.9,79.1h41.6c39.6,0,57,28.3,57,53.6c0,27.5-21.5,53.6-56.8,53.6h-41.8V79.1z M124.3,172.4h24.5c34.9,0,42.9-26.5,42.9-39.7c0-21.5-13.7-39.7-43.7-39.7h-23.7V172.4z}
                 svg{M88.7,56.8c0,5.5-4.5,10.1-10.1,10.1c-5.6,0-10.1-4.6-10.1-10.1c0-5.6,4.5-10.1,10.1-10.1C84.2,46.7,88.7,51.3,88.7,56.8z};
  }
}

\newcommand\orcidicon[1]{\href{https://orcid.org/#1}{\mbox{\scalerel*{
\begin{tikzpicture}[yscale=-1,transform shape]
\pic{orcidlogo};
\end{tikzpicture}
}{|}}}}

\def\BibTeX{{\rm B\kern-.05em{\sc i\kern-.025em b}\kern-.08em
T\kern-.1667em\lower.7ex\hbox{E}\kern-.125emX}}

\DeclareAcronym{PET}{ 
short = PET,
long = positron emission tomography
}

\DeclareAcronym{CLB}{ 
short = CLB,
long = cesium lead bromide
}

\DeclareAcronym{FBK}{ 
short = FBK,
long = Fondazione Bruno Kessler 
}

\DeclareAcronym{TOF}{ 
short = TOF,
long = time-of-flight 
}

\DeclareAcronym{CTR}{ 
short = CTR,
long = coincidence time resolution 
}

\DeclareAcronym{FWHM}{ 
short = FWHM,
long = full width at half maximum 
}

\DeclareAcronym{LOR}{ 
short = LOR,
long = line of response 
}

\DeclareAcronym{SNR}{ 
short = SNR,
long = signal-to-noise ratio
}

\DeclareAcronym{Seff}{ 
short = $S_{eff}$,
long = effective sensitivity 
}

\DeclareAcronym{LTE}{ 
short = LTE,
long = light transfer/collection efficiency
}

\DeclareAcronym{APTP}{ 
short = APTP,
long = avalanche photon transfer probability
}

\DeclareAcronym{PTS}{ 
short = PTS,
long = photon time spread 
}
    
    \DeclareAcronym{DOI}{ 
    short = DOI,
    long = depth-of-interaction
    }

       \DeclareAcronym{DOIs}{ 
    short = DOIs,
    long = depth-of-interactions,
    first-style = short
    }
    
    \DeclareAcronym{SiPM}{
      short = SiPM,
      long = silicon photomultiplier,
      first-style = short
      }
    
\DeclareAcronym{SiPMs}{
  short = SiPMs,
  long = silicon photomultipliers
}

\DeclareAcronym{SE}{
  short = SE,
  long = single-ended
}

\DeclareAcronym{DE}{
  short = DE,
  long = dual-ended
}

\DeclareAcronym{SPAD}{ 
short = SPAD,
long = single photon avalanche diode
}

\DeclareAcronym{SPADs}{ 
short = SPADs,
long = single photon avalanche diodes
}

\DeclareAcronym{SPTR}{ 
short = SPTR,
long = single photon time resolution
}

\DeclareAcronym{HAR}{ 
short = HAR,
long = high-aspect-ratio
}

\DeclareAcronym{BGO}{ 
short = BGO,
long = bismuth germanium oxide
}

\DeclareAcronym{MT}{ 
short = MT,
long = metal-trenched
}

\DeclareAcronym{LYSO}{ 
short = LYSO:Ce,
long = cerium doped lutetium–yttrium oxyorthosilicate
}

\DeclareAcronym{LNHF}{ 
short = LNHF,
long = low-noise and high-frequency
}

\DeclareAcronym{RF}{ 
short = RF,
long = radio frequency
}

\DeclareAcronym{BL}{ 
short = BL,
long = base line
}

\DeclareAcronym{GE}{ 
short = GE,
long = Gaussian equivalent
}

\DeclareAcronym{PDE}{ 
short = PDE,
long = photon detection efficiency
}

\DeclareAcronym{PCB}{ 
short = PCB,
long = printed circuit boards
}

\DeclareAcronym{PMTs}{ 
short = PMTs,
long = photomuliplier tubes
}

\DeclareAcronym{ER}{ 
short = ER,
long = energy resolution
}

\DeclareAcronym{MPPC}{ 
short = MPPC,
long = multi-pixel photon counter
}

\DeclareAcronym{LIDAR}{ 
short = LIDAR,
long = light detection and ranging
}

\DeclareAcronym{DCR}{ 
short = DCR,
long = dark count rate
}
\begin{document}

\bibliographystyle{ieeetr} 
\title{Light-Based Fast Timing in Bulk CsPbBr$_3$ Crystals for TOF-PET and Proton Range Verification}

\author{Nicolaus Kratochwil\orcidicon{0000-0001-5297-1878} \IEEEmembership{Member, IEEE}, Leonor Rebolo, Indra R. Pandey, Joshua W. Cates\orcidicon{0000-0002-5649-8691} \IEEEmembership{Member, IEEE},\\  Emilie Roncali\orcidicon{0000-0002-2439-1064} \IEEEmembership{Senior member,~IEEE}, Joshua H. Palmer, Gerard Ari\~no-Estrada\orcidicon{0000-0002-6411-191X} \IEEEmembership{Member, IEEE} \thanks{
%
This work was supported by the National Institutes of Health through grants R01 EB029533 and R01 EB034062 (PI Ari\~no-Estrada). L. Rebolo was supported by Prototera (PRT/BD/154945/2023)  \\ \textit{Corresponding authors: Nicolaus Kratochwil (nkratochwil@ucdavis.edu) and Gerard Ari\~no-Estrada (garino@ucdavis.edu)}.\\
This work did not involve human subjects or animals in its research.\\ 
Nicolaus Kratochwil is with the Department of Biomedical Engineering, University of California at Davis (UCD), Davis, CA, United States.
Leonor Rebolo is with the Department of Biomedical Engineering at UCD and with
I3N-Physics
Department, University of Aveiro, Portugal.
Indra R. Pandey and Joshua H. Palmer are with Actinia, Inc, Wilmette, IL, United States.
Joshua W. Cates is with the Lawrence Berkeley National Laboratory, Berkeley, CA, United States.
Emilie Roncali is with the Department of Biomedical Engineering at UCD.
Gerard Ari\~no-Estrada is with the Department of Biomedical Engineering at UCD and with the Institut de Física d'Altes Energies - Barcelona Institute of Science and Technology, Bellaterra, Barcelona, Spain. \\
%
}
}

\maketitle
\begin{abstract}
Halide perovskite semiconductors such as CsPbBr$_3$ (CLB) are emerging gamma-ray detectors for applications requiring very high energy resolution and potential for fine detector segmentation.
Semiconductor detectors typically offer poor time resolution due to the long drift times.
Recently, we proposed to use the Cherenkov light component in partially transparent semiconductors to boost the timing capability of such detectors.
Cherenkov light produced upon 511~keV gamma-ray interaction with CLB was investigated by means of optical simulations and experimental measurements.
%
%
The timing capability of a pair of identical CLB crystals ($\approx$ 3 x 3 x 5~mm$^3$) coupled to NUV-MT silicon photomultipliers was measured.
On average, 9.5 Cherenkov photons are produced in CLB between 555 and 900~nm for 511~keV photoelectric interactions based on our simulation framework.
Experimentally, we observe 2-to-3 times more photons detected than in the simulation. 
The two most likely explanations for these additional detected optical photons are either the partial transparency of CLB in the UV, or a mild scintillation light emitted by CLB at room temperature.
A coincidence time resolution (CTR) of 419~ps FWHM was obtained by triggering on more than 2 fired SiPM cells and after time walk correction.
%
%
%
The measured CTR confirms the feasibility to use the Cherenkov light-component for fast timing applications on top of the charge readout, toward full 3D localization.

\end{abstract}


\section{Introduction} 
\label{Introduction}

%

\IEEEPARstart{L}{ead} halide perovskite semiconductor materials have emerged relatively recently as a promising choice for gamma-ray radiation detectors that require very high energy resolution (ER) and detector granularity.
In particular, CsPbBr$_3$ (CLB) is attracting interest based on its high detection efficiency, the possibility to operate at room temperature, and relatively low production costs driven by its high growth yield~\cite{Stoumpos_ACS_2013,Liu_ACS_2022}.
ER as low as 1.4~\% at 662~keV has been reported with CLB crystals with 2.5 x 2.0 x 1.3~mm$^3$ crystal size ~\cite{He_2021_NP}. 

Time resolution has been the major drawback of semiconductor materials for gamma-ray detection, especially when they are compared against light-based scintillation detectors. 
This is due, in part, to the long drift times of their charge carriers, typically on the order of microseconds, one or two orders longer than the emission and transport of scintillation photons.
The asymmetry between the mobilities of electrons and holes in most high-Z materials also plays a role by introducing an additional time jitter depending on the interaction location of the gamma.
CLB is almost unique in that the mobilities of electrons and holes are very similar, 63 and 49~cm$^2$/V, respectively~\cite{He-2019-NIMA}.

Cherenkov light generated in partially transparent semiconductor materials has been recently proposed as an alternative to the charge induction mechanism to improve the time resolution in such detectors \cite{Arino-Estrada_2025_TRPMS}.
The extraction of fast timing signals using Cherenkov light has been reported in TlBr and TlCl for the detection of 511~keV gamma photons 
\cite{Arino-Estrada_2018_PMB,Arino-Estrada_2021_TRPMS,Terragni_2022_FrontPhysics,Kratochwil_TRPMS_2024,Mariscal-TRPMS_2024} and multi-MeV prompt-gammas \cite{Ellin_PMB_2024}.
Simultaneous measurements of Cherenkov light and charge induction signals have been reported with TlBr Cherenkov Charge Induction (CCI) detectors for 511~keV \cite{Arino-Estrada_2019_PMB} and 1.275~MeV gammas \cite{Arino-Estrada_2021_PMB}. 
Such reports show the capacity of obtaining a time resolution of few hundred picoseconds from the Cherenkov component, energy and 2D segmentation from the charge, and depth-of-interaction information by combining them, thus obtaining effective 3D information. 

CLB features partial transparency with a bandgap energy of 2.3 (transparent above 555~nm~\cite{Stoumpos_ACS_2013,Tao_2022_TRPMS,Arino-Estrada_2025_TRPMS}) and a high effective atomic number, making it a good candidate material for CCI detectors.
The refractive index of 2.4 is much higher than most scintillators, favoring the production of Cherenkov photons.
One study reports the extraction of fast signal from CLB using its light component \cite{Tao_2022_TRPMS}, reaching 440~ps full width at half maximum (FWHM) coincidence time resolution between a CLB crystal of 3$\times$3$\times$3~mm$^3$ operated in coincidence with a fast reference detector.
More recently, a CCI detector based on CLB has been proposed~\cite{Ellin_2024_IEEE}, following the analogous work reported with TlBr. 

In this contribution, we aim to deepen the understanding of Cherenkov light-based radiation component in CLB for the detection of gamma photons.
We first focus on the Cherenkov light yield.
We measure experimentally the number of detected photons with our setup and compare it against a simulation framework that we have used for similar studies previously \cite{Rebolo-TRPMS-2024}.
While the experimental measurements allow us to quantify the number of available Cherenkov photons with current SiPM technology, the simulation framework predicts how many Cherenkov photons are generated in the crystal for 511~keV photopeak interactions.
In the second part, we focus on the CTR between two CLB crystals with 511~keV gamma photons using exclusively experimental data.

The quantification of the Cherenkov light yield and achievable time resolution aims to help assess the potential of CLB semiconductor crystals as detectors in time-of-flight positron emission tomography (TOF-PET) and proton range verification (PRV) in proton therapy.

\section{Materials and Methods}

\subsection{CsPbBr$_3$ single crystals}
\begin{figure}[h!]
\begin{center}
\includegraphics[width=0.8\textwidth]{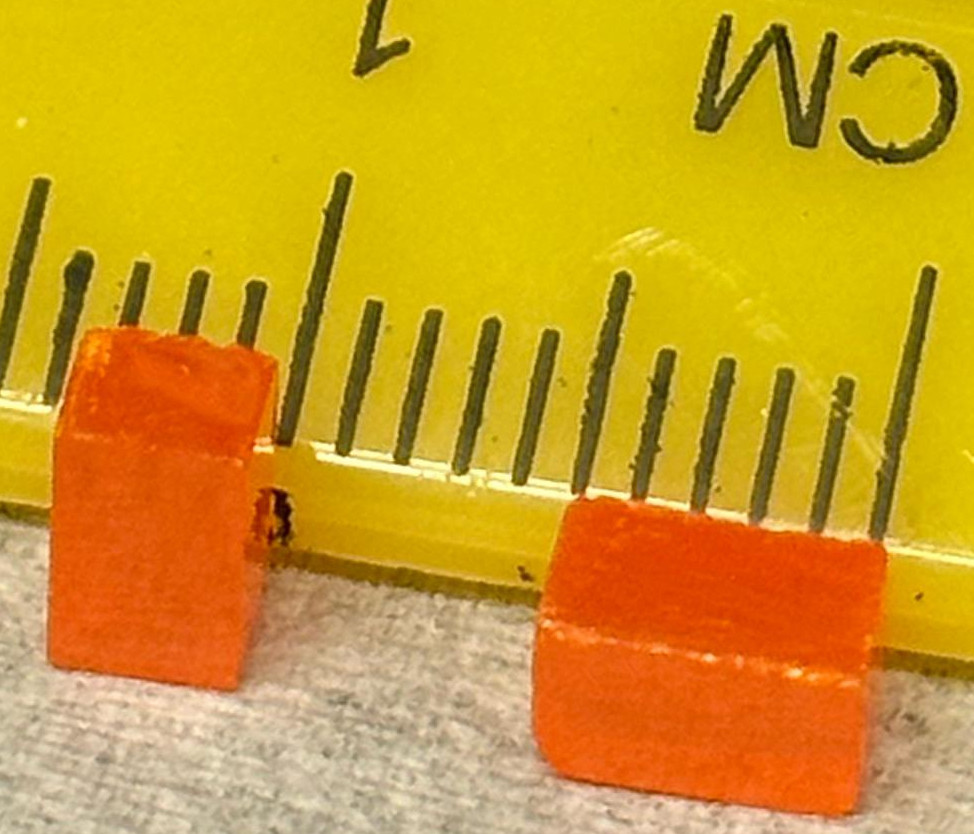}
\end{center}
\caption{3.2$\times$3.4$\times$5.4~mm$^3$ crystals employed in this work.}
\label{picture-CLB}
\end{figure}

CsPbBr$_3$ was synthesized by heating high-purity CsBr and PbBr$_2$ in a box furnace at 650~$^\circ$C, then crystallized using a Bridgman furnace, as detailed in previous publication~\cite{He_2021_NP}. 
To mitigate internal stress from phase transitions~\cite{Rodova-2003-JTAC,Chung_2024-crystal}, the crystal was cooled to room temperature in a two-step process with a 6 hour hold at 130~$^\circ$C.
The resulting crack-free single crystal was cut and manually polished with 600 and 1200 grade sandpaper, respectively.
Pictures of the crystals used in this study are shown on in figure~\ref{picture-CLB}.

\subsection{Experimental setup}

Measurements were conducted in a light-tight box at stable temperature ($\approx$~20~C$^\circ$).
Two CLB crystals were wrapped in Teflon and coupled with Cargille Meltmount (n=1.58) to NUV-MT SiPMs from Broadcom (AFBR-S4N44P014M).
A $^{22}$Na radioactive source with an activity of 6.3~MBq was placed between the two detectors to generate 511~keV gamma-rays in coincidence.
SiPMs were biased at 40~V, which corresponds to an overvoltage of around 8~V. 
The voltage drop of the SiPMs was read out with custom high-frequency readout electronics (based on the design of~\cite{Cates_2018_PMB,Cates_2022_PMB}) and the signal digitized with a Tektronix oscilloscope (4 GHz, 20~Gs/s). 
The raw waveforms were analyzed offline, in particular the signal amplitude (number of fired SiPM cells in each detector) and the crossing time for different leading edge thresholds after baseline correction~\cite{Kratochwil-2025-TRPMS}.
The trigger was set to detect coincidence events with more than 2 fired cells in each SiPM. 
This relatively high threshold was necessary to discard the majority of random coincidences coming from dark counts with secondary crosstalk events. 

\subsection{Data analysis}

The maximum signal  amplitude was extracted within 4~ns after passing the trigger threshold. 
Such a narrow time window avoided cross-contamination coming from delayed correlated noise~\cite{ACERBI_NIMA_2019} (SiPM afterpulses) with secondary crosstalk.

The coincidence time delay distribution was modeled either with a constant (for random coincidences from dark counts) plus one Gaussian distribution, or with the sum of a constant and two Gaussian distributions, similar to models reported for Cherenkov and scintillation emitters~\cite{Gundacker_2019_PMB,Kratochwil_2020_PMB}. 
The FWHM was calculated as 2.355 standard deviations (for the Gaussian + constant term) of the Gaussian distribution, or as the width of the two Gaussian at half the maximum (subtracting the constant term to compute the maximum and the half maximum, respectively).
A time walk correction was applied based on the signal amplitude (number of triggered cells) analog to~\cite{Kratochwil_2021_PMB,Mariscal-TRPMS_2024}. 
Within the precision of the analysis, identical CTR values were achieved with a correction based on the signal rise time instead.
%

The number of detected photons in CLB is expected to be very low, 5 photons per 511~keV gamma-ray interaction or less. 
Therefore, the classic comparison of pulse height spectra is not feasible to determine the light yield.
Instead, a statistical approach based on counting the number of triggered SiPM single photon avalanche diodes (SPADs) was used following the framework of~\cite{Arino-Estrada_2021_TRPMS,Kratochwil_2021_PMB}.

For a Poisson-distributed number of photons with a mean $\mu$ and a branching Poisson process to describe the SiPM crosstalk, the total event distribution of the number of triggered SiPM cells can be described according to equation~\ref{generalized-poisson}~\cite{VINOGRADOV_NIMA_2012}.

\begin{equation}
\label{generalized-poisson}
    P(N=k) = \frac{\mu \cdot (\mu + \lambda \cdot k)^{k-1} \cdot \text{exp}(-\mu - k \cdot \lambda) }{k!}
\end{equation}

The crosstalk contribution $\lambda$ was determined with a dark count measurement at the same SiPM bias voltage and the crystal coupled to the SiPM, but without radioactive source. 
The procedure is described in~\cite{Kratochwil_2025_sensors}.
The experimental event distribution $N(k) = N_0 \cdot P(k)$ was therefore fitted with 2 unknown parameters, namely $N_0$ and $\mu$.

\subsection{Optical simulations}

The toolkit GATE v9.0 (Geant4 10.06.03 and ROOT 6.20) was used to simulate the radiation-matter interactions between gamma rays and the CLB crystal block as described in~\cite{Rebolo-TRPMS-2024}.
A CLB crystal block of~3.2$\times$3.4$\times$5.4~mm$^3$ was simulated.
The photodetector was modeled with a 0.1 mm thick layer of silicon dioxide attached to the back face of the crystal.
A gamma source was placed at 32.3~mm from the front face of the crystal, modeled as a point source with isotropic emission at 511~keV.

The optical transport was modeled with the LUT Davis model~\cite{Trigila-2021-MedPhys}.
The crystal-photodetector interface was defined as a polished surface coupled to the photodetector with optical grease, assuming 100~$\%$ transparency for Cherenkov photons with wavelengths longer than 555~mm shown in figure~\ref{fig-SiPM}.
The interface between the crystal and the optical system was defined as a polished surface wrapped in four layers of Teflon tape.

\subsection{Post-processing}

\begin{figure}[h!]
\begin{center}
\includegraphics[width=0.99\textwidth]{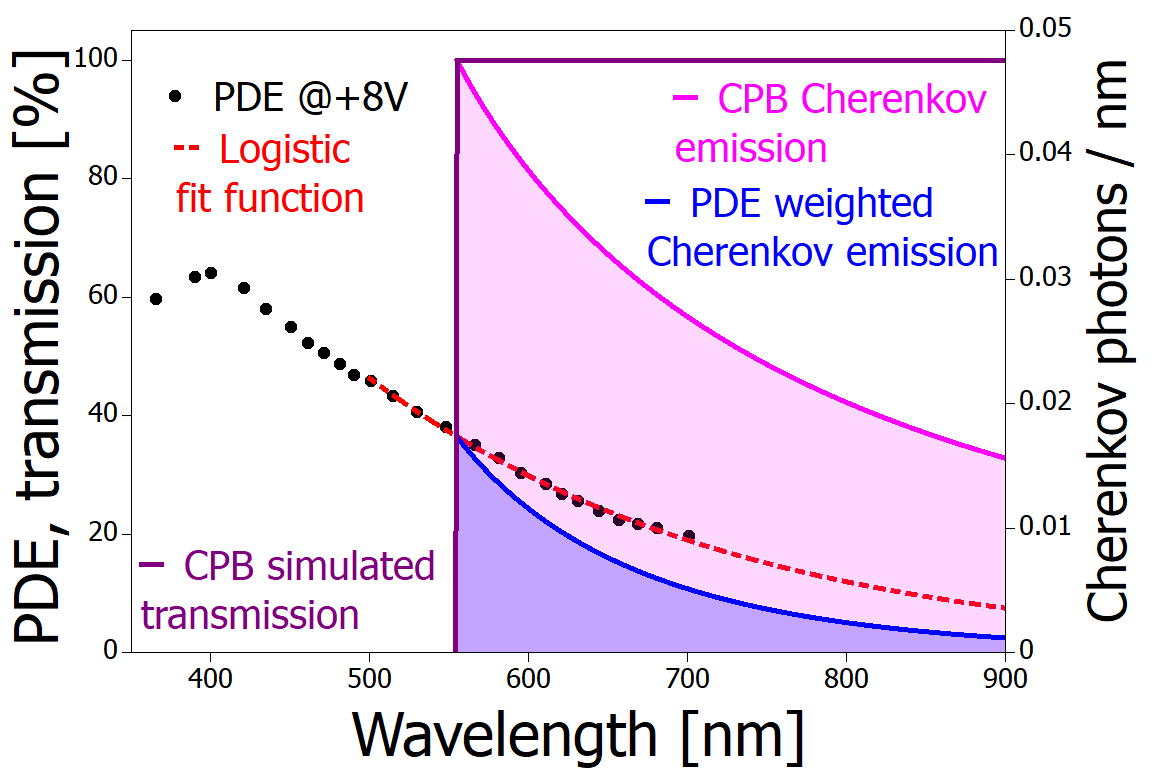}
\end{center}
\caption{Photon detection efficiency, simulated transparency, and calculated Cherenkov photon emission profile as function of the wavelength. The Cherenkov emission profile was calculated according to~\cite{Kratochwil_TRPMS_2024}.}
\label{fig-SiPM}
\end{figure}

A stochastic filter with a 21~\% photon detection efficiency was applied to the simulated photons arriving at the silicon dioxide layer to reproduce the SiPM photon detection efficiency (PDE).
This PDE value was derived by first approximating the reported PDE~\cite{Merzi_2023_JINST} beyond 700~nm with a logistic function and later convolving it with the Cherenkov emission of CLB for each wavelength.
The weighted detection efficiency is the ratio of the areas under the curve between the PDE times the Cherenkov emission and the Cherenkov emission, as shown in figure~\ref{fig-SiPM}.

SiPM crosstalk was added to the distribution using a branching Poisson process. 
Each detected photon may trigger an additional number of cells with the event distribution for one primary photon triggering $k$ cells described in equation~\ref{Borel}~\cite{VINOGRADOV_NIMA_2012}.

\begin{equation}
\label{Borel}
    P_{(1\rightarrow k)} = \frac{ (\lambda \cdot k)^{k-1} \cdot \text{exp}( - k \cdot \lambda) }{k!}
\end{equation}

The probability that N primary photons trigger M$\geq$N cells due to crosstalk was calculated iteratively for increasing N according to equation~\ref{iteratively}.

\begin{equation}
   \begin{gathered}
    P_{(N\rightarrow M)} = P_{(1\rightarrow 1)} \cdot P_{(N-1\rightarrow M-1)}  \\ + P_{(1\rightarrow 2)} \cdot P_{(N-1\rightarrow M-2)} + ... \\
    +   P_{(1\rightarrow M-N+1)} \cdot P_{(N-1\rightarrow N-1)} \\
    P_{(N\rightarrow M)} = \sum_{i=1}^{M-N+1} P_{(1\rightarrow i)} \cdot P_{(N-1 \rightarrow M-i)}
    \label{iteratively}
    \end{gathered}
\end{equation}

\section{Results}
\subsection{Simulated Cherenkov photons in CLB}
\label{sec-simulated}

\begin{figure}[h!]
\begin{center}
\includegraphics[width=0.99\textwidth]{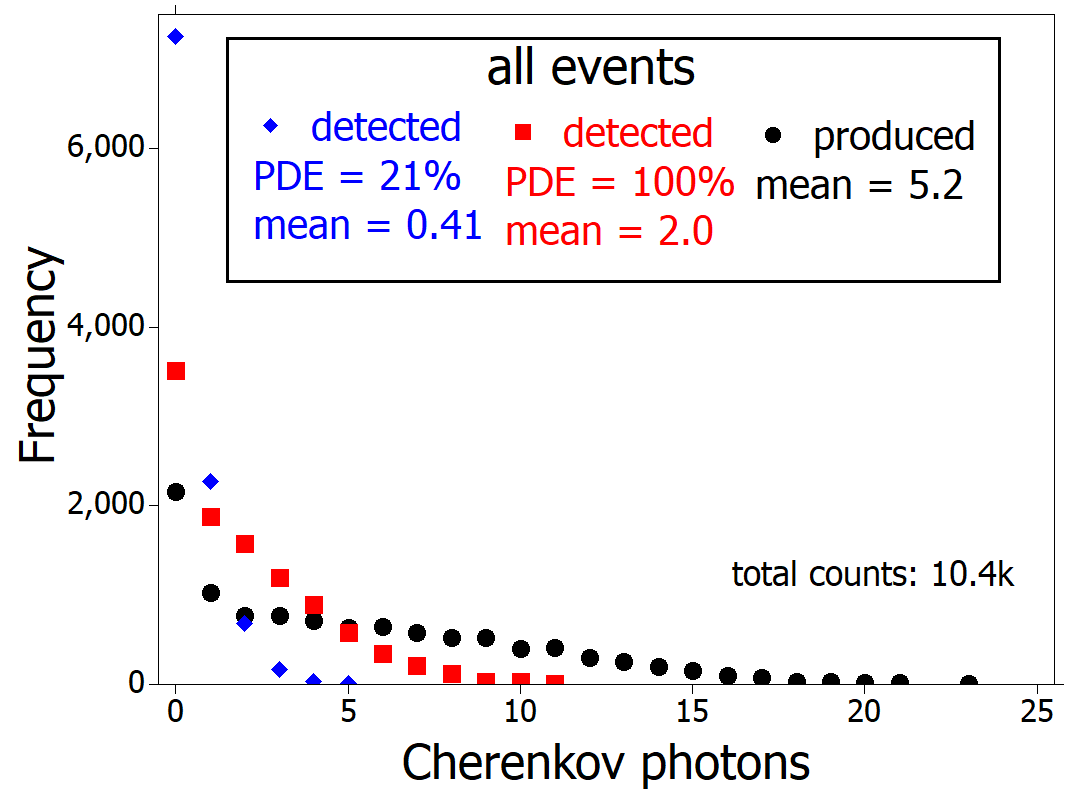} \\
\includegraphics[width=0.99\textwidth]{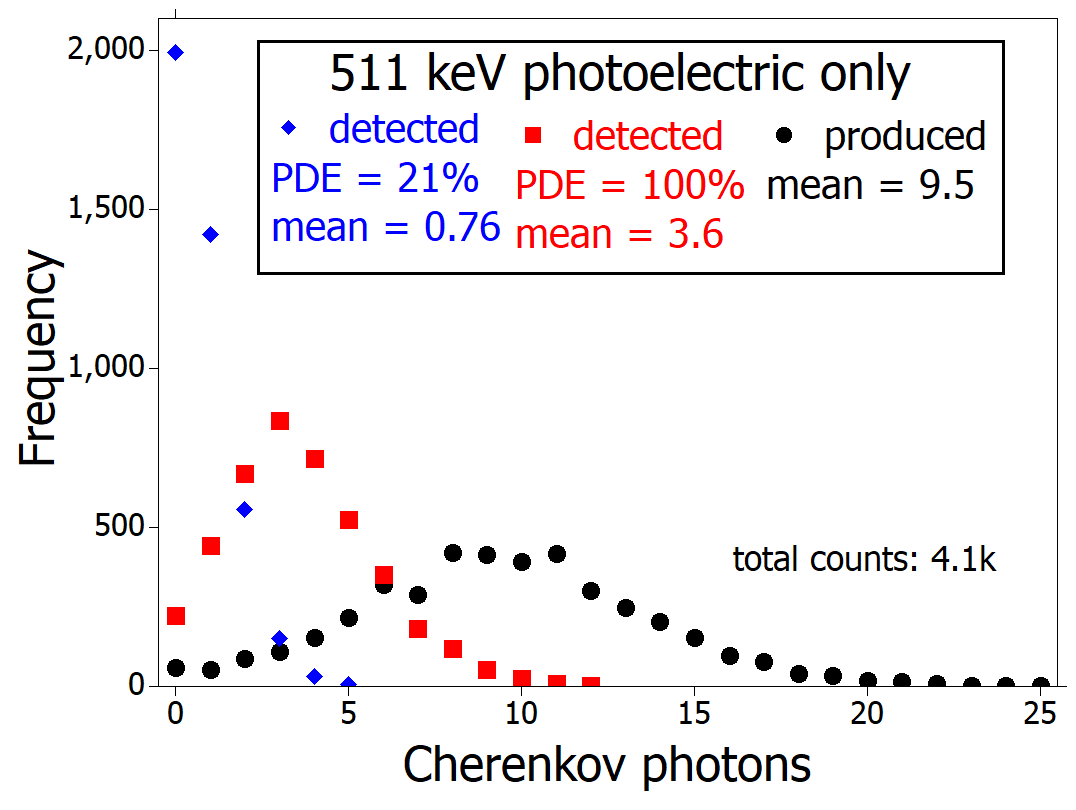} 
\end{center}
\caption{Simulated Cherenkov photons in CLB considering all interactions (top) or 511~keV photoelectric interactions only (bottom). }
\label{simulation-results}
\end{figure}

The simulated number of produced Cherenkov photons, detected with perfect PDE, and detected with realistic PDE is shown in figure~\ref{simulation-results} either for all events (top) or upon selecting on 511~keV photoelectric interactions (bottom). 
Less than 40\% of the produced photons reach the photodetector likely due to the large mismatch of the refractive indices between crystal and optical coupling. 
The mean number of produced or detected photons is almost by a factor 2 lower considering all events.
This is expected as the majority of events (64\%) interact via Compton interaction and there is a strong non-linearity on the number of produced Cherenkov photons at low energies.

\subsection{Experimental photon yield in CLB}

\begin{figure}[h!]
\begin{center}
\includegraphics[width=0.99\textwidth]{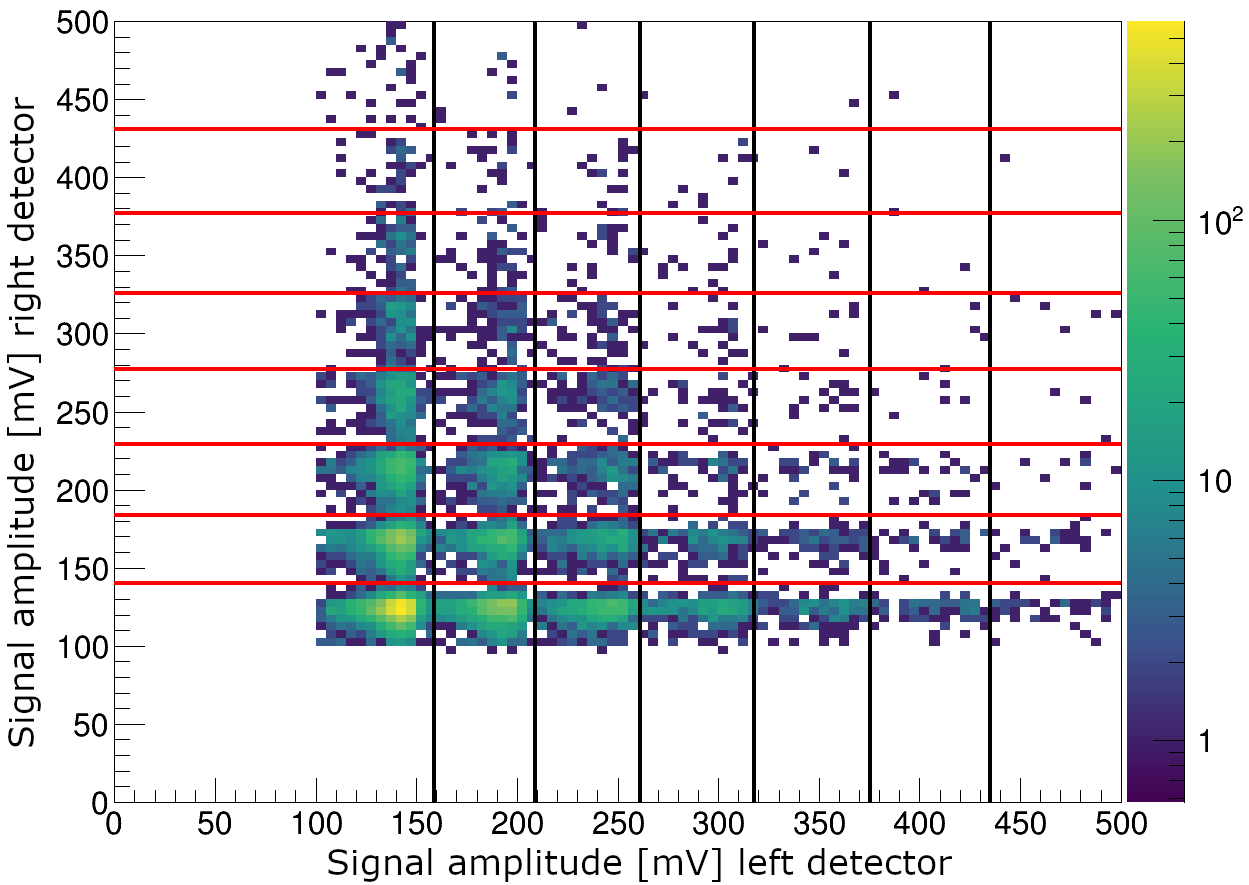} \\
\includegraphics[width=0.99\textwidth]{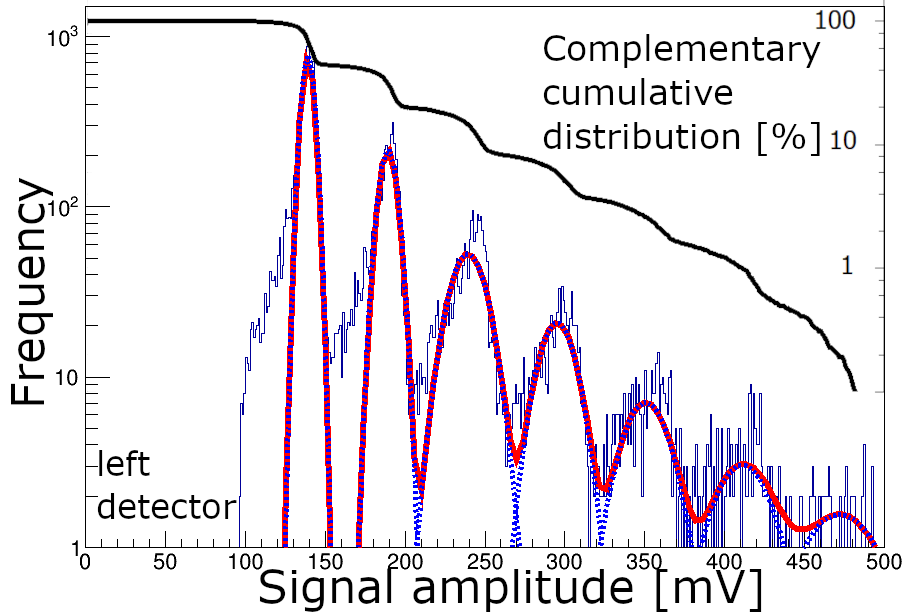}
\end{center}
\caption{Top: 2D histogram of the signal amplitude. The red and black lines serve as eye-guide to distinguish different number of triggered cells from 3 to 9. Bottom: Histogram of the signal amplitude for one detector. The distribution is fitted with a sum of 7 gaussian distributions (red) with each contribution drawn in blue. The complementary cumulative distribution (normalized) is shown in black with the right y-axis.}
\label{experiment-fingers}
\end{figure}

\begin{figure}[h!]
\begin{center}
\includegraphics[width=0.99\textwidth]{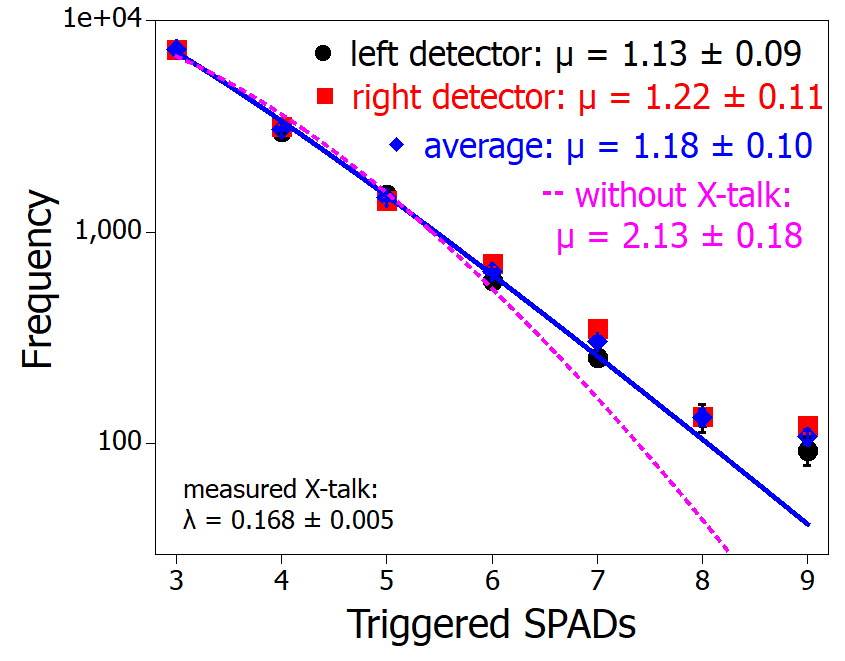} \\
\end{center}
\caption{Frequency of the triggered SPADs for the two detectors (black, red) and average (blue) with fit function considering measured crosstalk (blue solid line). Ignoring crosstalk, the mean number of detected photons is almost 2-fold overestimated.}
\label{yield-results}
\end{figure}

The figure~\ref{experiment-fingers}-top shows a 2D histogram of the number of triggered cells in coincidence. 
Only events within $\pm$ 500~ps were used to construct the fingerplots (bottom of figure~\ref{experiment-fingers}) of the two detectors. 
This selection allowed to reduce random coincidences originating from dark counts in one of the two detectors, which follow different probability distributions~\cite{Kratochwil_2021_PMB}.
The frequency of the triggered cells of the two detectors is shown in figure~\ref{yield-results}. 
A fit was performed for both detectors individually using the measured crosstalk contribution of $\lambda~=~0.168~\pm~0.005$ and equation~\ref{generalized-poisson}. 
The average number of detected photons amounts to $\mu = 1.18 \pm 0.10$. 

\subsection{Coincidence time resolution}

\begin{figure*}[h!]
\begin{center}
\includegraphics[width=0.45\textwidth]{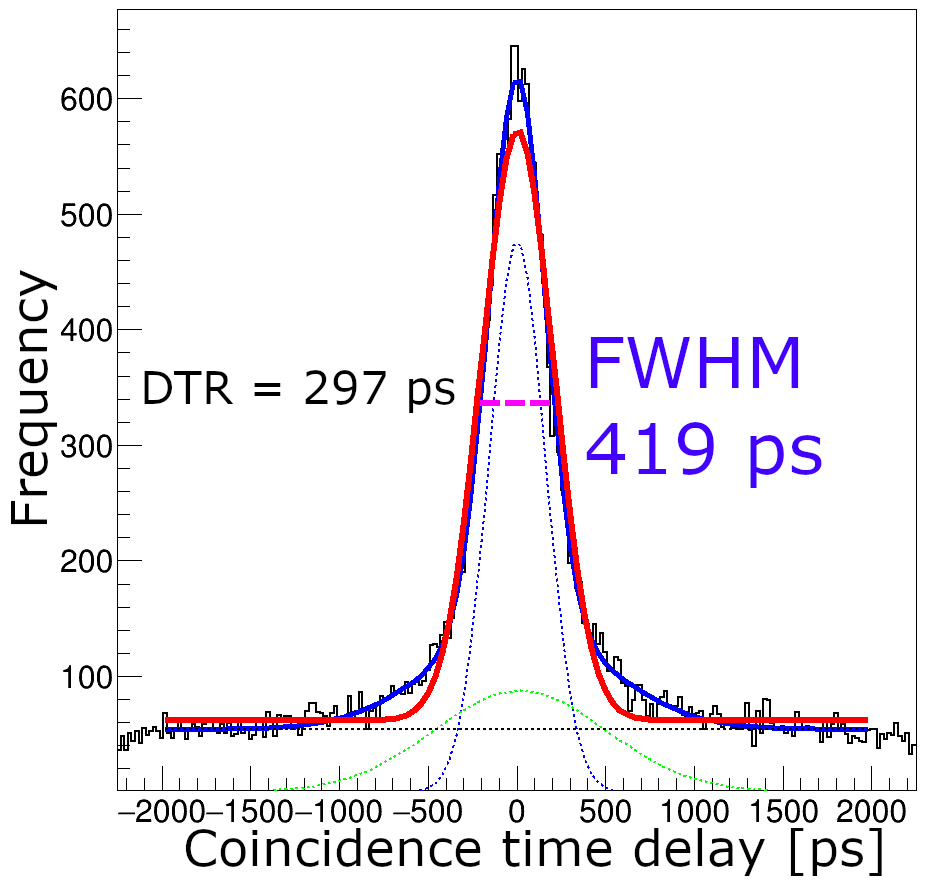}
\includegraphics[width=0.54\textwidth]{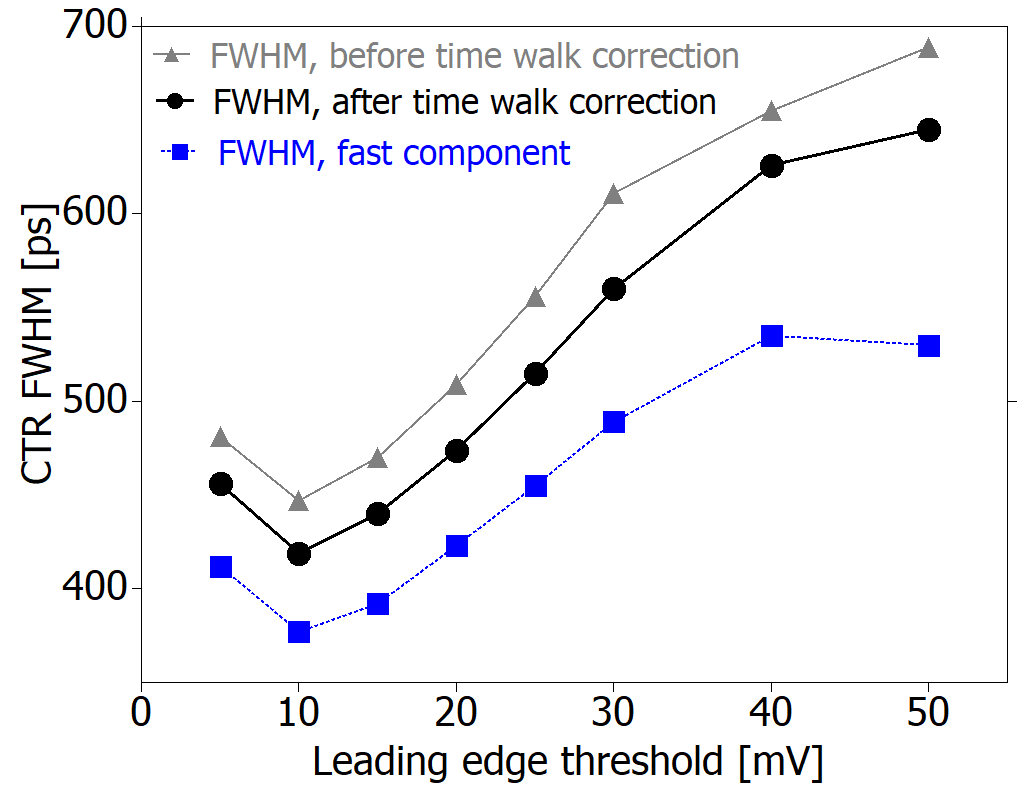}
\end{center}
\caption{Coincidence time resolution of two identical CLB detectors. Left: Histogram (black), fitted with a sum of two Gaussian fit functions (dotted blue and dotted green line, respectively) and a constant for the DCR (black dotted line). Right: Coincidence time resolution (FWHM) as function of the leading edge threshold. Points are connected as eye-guide.} 
\label{CTR}
\end{figure*}

The time delay histogram using the same dataset but a larger coincidence window is shown on the left of figure~\ref{CTR} after time walk correction.
A fit function consisting of a constant and a single Gaussian function was tested (in red), but does not represent well the measured time delay distribution.
A FWHM of 419~$\pm$~5~ps was achieved for two detectors in coincidence with a double-gaussian fit (blue). 
The single detector time resolution (DTR) is 297~$\pm$~4~ps, which is much better compared to the values reported in~\cite{Tao_2022_TRPMS} of $\approx$ 440~ps.
The time resolutions as function of the leading edge threshold before and after time walk correction as well as only the fast component are shown in figure~\ref{CTR}, right. 
 
Tests at higher SiPM overvoltage were performed with the best CTR result of 345~$\pm$~15~ps FWHM (244~$\pm$~11~ps DTR) at 44~V bias voltage and triggering on 4 or more triggered cells. 
However, the higher dark count rate makes the modeling of the time delay distribution more prone to instabilities. 

\section{Discussion}

\begin{figure}[h!]
\begin{center}
\includegraphics[width=0.95\textwidth]{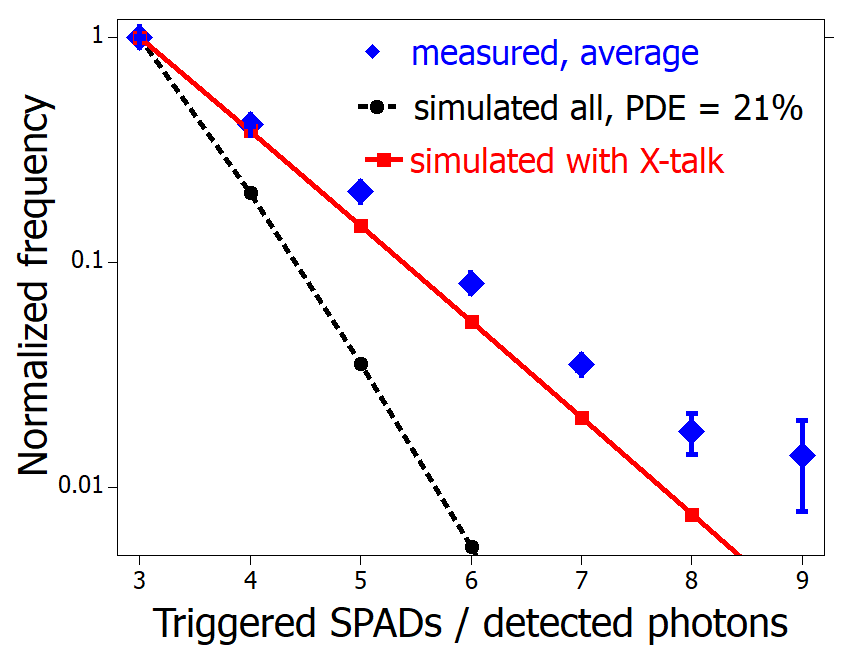}
\end{center}
\caption{Normalized frequency of the number of triggered cells / detected photons of the measured (blue) and simulated data considering all events without (black) and with (red) SiPM crosstalk.}
\label{comparison}
\end{figure}

On average, we detected 1.18 $\pm$ 0.10 optical photons. 
This value is almost 3 times larger than the simulated Cherenkov photon yield of 0.41$\pm$ 0.01 for all events. 
However, in the simulation we used the average of the distribution without any fitting.
For the experimental results, taking the average was not possible due to SiPM crosstalk contamination and the lack of events with 0, 1, or 2 triggered SPADs.
Therefore, the mean value was extracted based on a fit function with the underlying model that the number of detected photons is Poisson distributed.
%
This assumption is not fully supported by the simulated dataset considering all events.
The reason is that the average among different Poisson distributions (corresponding to different energy depositions and therefore different mean number of produced Cherenkov photons) does not follow a Poisson distribution.
To mitigate any potential bias from fitting the distribution with a non-ideal model, we find it more appropriate to directly compare the two datasets.
We account for SiPM crosstalk based on equations~\ref{Borel} and \ref{iteratively} in figure~\ref{comparison}.

This scenario also suggests that more photons are detected in the experimental event distribution than those predicted by simulations. 
We discuss two possible explanations for this additional number of detected optical photons: 
First,  the transmission measurement reported in~\cite{Tao_2022_TRPMS} shows that not all photons are absorbed below the cutoff wavelength.
Since the Cherenkov emission follows a 1/$\lambda^2$ distribution and the SiPM PDE is also higher in the UV, even a few escaping UV Cherenkov photons can alter the distribution. 

A second explanation is that CLB is known to scintillate at cryogenic temperatures with reported light yield values between 50 and 109 ph/keV at 7~K~\cite{Mykhaylyk-2020-SR} and fast decay times~\cite{Nikl-1994-CPL}. 
For increasing temperature (reported up to 170~K) the scintillation is highly quenched.
Nevertheless, even a very low scintillation of around 17 photons per MeV is sufficient to explain the difference.
While there exists no literature about scintillation in bulk CLB at room temperature, for nano-sized CLB the scintillation emission can be ultra-fast with decay times similar to Cherenkov emission~\cite{Decka_2022_JMaterChem}.

The reported time resolution of 419~$\pm$ 5~ps FWHM does not account for any energy selection for events with at least three triggered cells.
For monenergetic  photoelectric energy deposition at 511~keV, the mean number of detected Cherenkov photons is higher, hence a better CTR can be anticipated with photopeak discrimination.
Further, the  (NUV-MT) SiPMs used in this experiment were not optimized for this crystals and red-sensitive SiPMs will have a higher detection efficiency for CLB.
The high SiPM dark count rate at room temperature was a challenge to get true coincidence events.
For proton range verification the gamma rays have energies between 1 to 7~MeV, which is up to one order of magnitude larger energy deposition, resulting in much more detected photons. 
This would allow to both increase the SiPM bias voltage (with proven CTR improvement) as well as placing the trigger threshold at a higher level to largely eliminate random dark counts.

\section{Conclusions}

Light emission in bulk CsPbBr$_3$ semiconductor crystals exposed to gamma-rays has been studied.
Differences between simulation and experimental results suggest there is an additional light component to Cherenkov light emitted above the cutoff wavelength of 555~nm.
The source of such light component is subject to further investigation.

%

The measured coincidence time resolution of 419~ps FWHM (297~ps DTR) is about 200~ps better to literature and can be used to provide an accurate time stamp for TOF-PET and PRV.
%
%
%
%
These results add to the enthusiasm raised by the reported performance of CLB detectors as semiconductor detectors and make it an attractive candidate for CCI gamma-ray detectors.

\section*{Acknowledgment}

The dataset used and or analyzed during the current study are available from the corresponding authors on reasonable request. \\
All authors declare that they have no known conflicts of interest in terms of competing financial interests or personal relationships that could have an influence or are relevant to the work reported in this paper.\\
All authors gave their approval for the final version of the manuscript.

\newpage
\bibliography{references_Sept2024}

\end{document}